\documentclass[conference]{IEEEtran}

\usepackage[final]{graphicx}
\usepackage{tabls}
\usepackage{amsmath,amssymb,amscd,latexsym,dsfont,amsthm}
\usepackage{cite}
\usepackage{psfrag}
\usepackage{multicol}
\usepackage{color,comment}
\usepackage{balance}
\usepackage{color}
\usepackage[ruled,vlined]{algorithm2e}
\usepackage{balance}

\IEEEoverridecommandlockouts
\begin{document}
\title{On the Secrecy Capacity of the Broadcast\vspace{0.2cm}\\ Wiretap Channel with Limited CSI Feedback}

\author{Amal Hyadi, Zouheir Rezki, and Mohamed-Slim Alouini \\Computer, Electrical, and Mathematical Sciences \& Engineering (CEMSE) Division
\\King Abdullah University of Science and Technology (KAUST), Thuwal, Makkah Province, Saudi Arabia\\ $\lbrace$amal.hyadi, zouheir.rezki, slim.alouini$\rbrace$@kaust.edu.sa
\thanks{
This work was supported by the Qatar National Research Fund (a member of Qatar Foundation) under NPRP Grant NPRP 5-603-2-243. The statements made herein are solely the responsibility of the authors. 
}
}


\maketitle
\normalsize

\begin{abstract}
In this paper, we investigate the problem of secure broadcasting over block-fading channels with limited channel knowledge at the transmitter. More particularly, we analyze the effect of having imperfect channel state information (CSI) via a finite rate feedback on the throughput of a broadcast channel where the transmission is intended for multiple legitimate receivers in the presence of an eavesdropper. Indeed, we consider that the transmitter is only provided by a $b$-bits feedback of the main channel state information (CSI). The feedback bits are sent by each legitimate receiver, at the beginning of each fading block, over error-free public links with limited capacity. Also, we assume that the transmitter is aware of the statistics of the eavesdropper's CSI but not of its channel's realizations. Assuming an average transmit power constraint, we establish upper and lower bounds on the ergodic secrecy capacity. We consider both the common message transmission case, where the source broadcasts the same information to all the receivers, and the independent messages case, where the transmitter broadcasts multiple independent messages to the legitimate receivers. In both scenarios, we show that the proposed lower and upper bounds coincide asymptotically as the capacity of the feedback links become large, i.e.~$b\rightarrow\infty$; hence, fully characterizing the secrecy capacity in this case.
\end{abstract}

\section{Introduction}
One of the most prominent concerns of wireless communication systems is to ensure the security of the users against eavesdropping attacks. In last years, wireless physical layer security has gained a lot of attention from the research community. Compared to the traditional cryptographic techniques, physical layer security is performed with less complexity and is more convenient for the emerging ad-hoc and decentralized networks and the next wave of innovative systems known as the Internet of Things. The information theoretic security was firstly introduced by Shannon in \cite{shannon01}. Many years later, Wyner proposed a new model for the wiretap channel \cite{wyner01}. In his seminal work, Wyner presented the degraded wiretap channel, where a source exploits the structure of the medium channel to transmit a message reliably to the intended receiver while leaking asymptotically no information to the eavesdropper. Ulterior works generalized Wyner's work to the case of non-degraded channels \cite{csiszar01}, Gaussian channels \cite{leung11}, and fading channels \cite{gopala01,liang01}. 

Recent research interest has been given to analyzing the secrecy capacity of multi-users systems. For the broadcast multi-users scenario, the secrecy capacity of parallel and fading channels, assuming perfect main CSI at the transmitter, was considered in~\cite{ashish01} and\cite{ulukus1}. The case of imperfect main channel estimation was elaborated in~\cite{hyadi2}. For the multiple access scenario, the authors in~\cite{tekin01} and \cite{ulukus3} investigated the secrecy capacity of degraded wiretap channels. The problem of analyzing the secrecy capacity of multiple antenna systems has also been of great interest. The secrecy capacity for the multiple-input single-output (MISO) wiretap Gaussian channel was proven in \cite{ashish04} and \cite{li2}. Another work \cite{ashish02} characterized the secrecy capacity for the MISO case, with a multiple-antenna eavesdropper, when the main and the eavesdropping channels are known to all terminals. The secrecy capacity of the multiple-input multiple-output (MIMO) transmission with a multiple-antenna eavesdropper was considered in \cite{oggier11} and \cite{ashish11} when the channel matrices are fixed and known to all terminals. The secrecy capacity region of the Gaussian MIMO broadcast wiretap channel was derived in \cite{ulukus2}.

Taking full advantage of the ability of the physical layer to secure wireless transmissions, requires a complete knowledge of the channel state information (CSI) at the transmitter (CSIT); which is difficult to have in practical scenarios. One way to overcome this challenge is by using feedback \cite{heath01,love11}. This is a natural setting as CSI feedback is incorporated in most if not all communication standards. For the case of a single user transmission, an upper and a lower bounds on the secrecy capacity were presented, in \cite{rezki11}, for block-fading channels with finite rate feedback. For the MIMO case, the work in \cite{lin11} and \cite{liu01} evaluated the impact of quantized channel information on the achievable secrecy rate, for multiple-antenna wiretap channels, using artificial noise.

In this paper, we investigate the ergodic secrecy capacity of a broadcast wiretap channel where a source transmits to multiple legitimate receivers in the presence of an eavesdropper. In particular, we analyze the case of block-fading channels with limited CSI feedback. More specifically, we consider that the transmitter is unaware of the channel gains to the legitimate receivers and to the eavesdropper and is only provided a finite CSI feedback. This feedback is sent by the legitimate receivers through error-free links with limited capacity. Both the common message transmission, where the same message is broadcasted to all the legitimate receivers, and the independent messages transmission, where the source broadcasts multiple independent messages, are considered. Assuming an average power constraint at the transmitter, we provide an upper and a lower bounds on the ergodic secrecy capacity for the common message case, and an upper and a lower bounds on the secrecy sum-rate for the independent messages case. For the particular case of infinite feedback, we prove that our bounds coincide. 

The paper is organized as follows. Section~\ref{model} describes the system model. The main results along with the corresponding proofs are introduced in section~\ref{BCM} for the common message transmission and in section~\ref{BIM} for the independent messages case. Finally, selected simulation results are presented in section~\ref{NR}, while section~\ref{conclusion} concludes the paper. 

\textit{Notations:}
Throughout the paper, we use the following notational conventions. The expectation operation is denoted by $\mathbb{E}[.]$, the modulus of a scalar $x$ is expressed as $|x|$, and we define $\{\nu\}^+{=}\max (0,\nu)$. The entropy of a discrete random variable $X$ is denoted by $H(X)$, and the mutual information between random variables $X$ and $Y$ is denoted by $I(X;Y)$. A sequence of length $n$ is denoted by $X^n$, $X(k)$ represents the $k$-th element of $X$, and $X(l,k)$ the $k$-th element of $X$ in the $l$-th fading block.
\section{System Model}\label{model}
\begin{figure}[t]
\psfrag{t}[l][l][1.9]{$\text{T}$}
\psfrag{e}[l][l][1.9]{$\text{E}$}
\psfrag{r1}[l][l][1.9]{$\text{R}_1$}
\psfrag{r2}[l][l][1.9]{$\text{R}_k$}
\psfrag{r3}[l][l][1.9]{$\text{R}_K$}
\psfrag{h1}[l][l][1.5]{$h_1$}
\psfrag{h2}[l][l][1.5]{$h_k$}
\psfrag{h3}[l][l][1.5]{$h_K$}
\psfrag{g}[l][l][1.5]{\hspace{-0.1cm}$h_\text{e}$}
\psfrag{feedback}[l][l][1.5]{\hspace{0.3cm}$b$-bits Feedback Link}
\psfrag{l}[l][l][2]{$\left. \rule{0pt}{2.2cm} \right\}$}
\psfrag{transmitter}[l][l][1.7]{Transmitter}
\psfrag{eavesdropper}[l][l][1.7]{Eavesdropper}
\psfrag{receivers}[l][l][1.7]{$\hspace{-0.1cm}K$ Legitimate Receivers}
\begin{center}
\scalebox{0.4}{\includegraphics{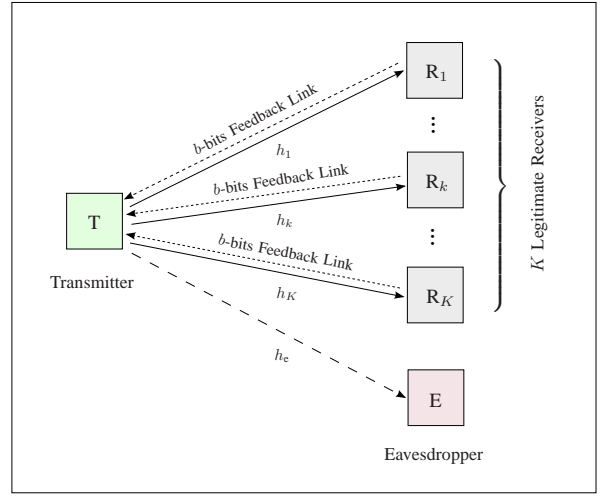}}
\end{center}
\caption{Broadcast wiretap channel with limited CSI feedback.}\vspace{-0.2cm}
\label{fig:model}
\end{figure}

We consider a broadcast wiretap channel where a transmitter~$\text{T}$ communicates with $K$ legitimate receivers $(\text{R}_1,\cdots,\text{R}_K)$ in the presence of an eavesdropper $\text{E}$ as depicted in Fig.~\ref{fig:model}. Each terminal is equipped with a single antenna for transmission and reception. The respective received signals at each legitimate receiver $\text{R}_k; k\in\{1,\cdots,K\},$ and the eavesdropper, at fading block $l$, $l{=}1,{\cdots ,}L$, are given by
\begin{equation}\label{Sys_Out}
\begin{aligned}
&Y_k(l,j)=h_k(l)X(l,j)+v_k(l,j)\\
&Y_\text{e}(l,j)\hspace{0.06cm}=h_\text{e}(l)X(l,j)\hspace{0.05cm}+w_\text{e}(l,j),
\end{aligned}
\end{equation}
where $j{=}1,{\cdots ,}\kappa$, with $\kappa$ representing the length of each fading block, $X(l,j)$ is the $j$-th transmitted symbol in the $l$-th fading block, $h_k(l)\!\in\!\mathbb{C}$, $h_\text{e}(l)\!\in\!\mathbb{C}$ are the complex Gaussian channel gains corresponding to each legitimate channel and the eavesdropper's channel, respectively, and $v_k(l,j)\!\in\!\mathbb{C}$, $w_\text{e}(l,j)\!\in\!\mathbb{C}$ represent zero-mean, unit-variance circularly symmetric white Gaussian noises at $\text{R}_k$ and~$\text{E}$, respectively. We consider a block-fading channel where the channel gains remain constant within a fading block, i.e., $h_k(\kappa l)=h_k(\kappa l-1)=\cdots =h_k(\kappa l-\kappa +1)=h_k(l)$ and $h_\text{e}(\kappa l)=h_\text{e}(\kappa l-1)=\cdots =h_\text{e}(\kappa l-\kappa +1)=h_\text{e}(l)$. We assume that the channel encoding and decoding frames span a large number of fading blocks, i.e., $L$ is large, and that the blocks change independently from a fading block to another. 
An average transmit power constraint is imposed at the transmitter such that 
\begin{equation}\label{pavg}
\frac{1}{n}\sum_{t=1}^n\mathbb{E}\left[|X(t)|^2\right]\leq P_{\text{avg}},
\end{equation}
with $n{=}\kappa L$, and where the expectation is over the input distribution.

The channel gains $h_k$ and $h_\text{e}$ are independent, ergodic and stationary with bounded and continuous probability density functions (PDFs). In the rest of this paper, we denote $|h_k|^2$ and $|h_\text{e}|^2$ by $\gamma_k$ and $\gamma_\text{e}$, respectively. We assume perfect CSI at the receiving nodes. That is, each legitimate receiver is instantaneously aware of its channel gain $h_k(l)$, and the eavesdropper knows $h_\text{e}(l)$, with $l{=}1,\cdots ,L$. The statistics of the main and the eavesdropping channels are available to all nodes. Further, we assume that the transmitter is not aware of the instantaneous channel realizations of neither channel. However, each legitimate receiver provides the transmitter with $b$-bits CSI feedback through an error-free orthogonal channel with limited capacity. This feedback is transmitted at the beginning of each fading block and is also tracked by the other legitimate receivers. The eavesdropper knows all channels and also track the feedback links so that they are not sources of secrecy.

The adopted feedback strategy consists on partitioning the main channel gain support into $Q$ intervals $[\tau_1, \tau_2), \cdots , [\tau_q, \tau_{q+1}), \cdots , [\tau_Q, \infty)$, where $Q{=}2^b$. That is, during each fading block, each legitimate receiver $\text{R}_k$ determines in which interval, $[\tau_q, \tau_{q+1})$ with $q{=}1,\cdots ,Q$, its  channel gain $\gamma_k$ lies and feedbacks the associated index $q$ to the transmitter. At the transmitter side, each feedbacked index $q$ corresponds to a power transmission strategy $P_q$ satisfying the average power constraint. We assume that all nodes are aware of the main channel gain partition intervals $[\tau_1, \tau_2), \cdots , [\tau_q, \tau_{q+1}), \cdots , [\tau_Q, \infty)$, and of the corresponding power transmission strategies $\{P_1, \cdots , P_Q\}$. 

The transmitter wishes to send some secret information to the legitimate receivers. In the case of common message transmission, a $(2^{n\mathcal{R}_\text{s}}, n)$ code consists of the following elements:
\begin{itemize}
\item A message set $\mathcal{W}=\left\lbrace 1,2,{\cdots},2^{n\mathcal{R}_\text{s}}\right\rbrace$ with the messages $W\in\mathcal{W}$ independent and uniformly distributed over $\mathcal{W}$;
\item A stochastic encoder $f: \mathcal{W}\rightarrow\mathcal{X}^n$ that maps each message $w$ to a codeword $x^n\in\mathcal{X}^n$;
\item A decoder at each legitimate receiver $g_k: \mathcal{Y}_k^n\rightarrow\mathcal{W}$ that maps a received sequence $y_k^n\in\mathcal{Y}_k^n$ to a message $\hat{w}_k\in\mathcal{W}$.
\end{itemize}

A rate $\mathcal{R}_\text{s}$ is an \textit{achievable secrecy rate} if there exists a sequence of $(2^{n\mathcal{R}_\text{s}}, n)$ code such that both the average error probability at each legitimate receiver
\begin{equation}
P_{\text{e}_k}=\frac{1}{2^{n\mathcal{R}_\text{s}}}\sum_{w=1}^{2^{n\mathcal{R}_\text{s}}}\text{Pr}\left[W\neq\hat{W}_k\big|W=w\right],
\end{equation} 
and the leakage rate at the eavesdropper
\begin{equation}
\frac{1}{n} I(W;Y_e^n,h_e^L,h_1^L,\cdots,h_K^L,q_1^L,\cdots,q_K^L),
\end{equation} 
go to zero as $n$ goes to infinity. The \textit{secrecy capacity} $\mathcal{C}_\text{s}$ is defined as the maximum achievable secrecy rate, i.e., $\displaystyle{\mathcal{C}_\text{s}\triangleq\sup\mathcal{R}_\text{s},}$ where the supremum is over all achievable secrecy rates. In the case of independent messages transmission, the achievable secrecy rate tuple $\left(\mathcal{R}_1,\mathcal{R}_2,\cdots,\mathcal{R}_K\right)$ can be analogously defined. 
\section{Broadcasting a Common Message}\label{BCM}
In this section, we examine the case of common message transmission when a unique confidential information is broadcasted to all the legitimate receivers in the presence of an eavesdropper. Taking into account the adopted system model, we present an upper and a lower bounds on the ergodic secrecy capacity.
\subsection{Main Results}\label{MR1}
\textit{Theorem 1:} The common message secrecy capacity, $\mathcal{C}_{\text{s}}$, of a block fading broadcast channel, with an error free $b$-bits CSI feedback sent by each legitimate receiver, at the beginning of each fading block, is bounded by
\begin{equation}\label{th1}
\mathcal{C}_{\text{s}}^-\leq \mathcal{C}_{\text{s}}\leq \mathcal{C}_{\text{s}}^+,
\end{equation}
such as\vspace{-0.3cm}
\begin{subequations}
\begin{align}\label{th1_a}
\mathcal{C}_{\text{s}}^-{=}\hspace{-0.2cm}\min_{1\leq k\leq K}\max_{\{\tau_q;P_q\}_{q=1}^Q}\sum_{q=1}^Q\Theta_{\tau_q}^{\gamma_k}\underset{\gamma_\text{e}}{\mathbb{E}}\!\left[\left\lbrace\log\!\left(\frac{1{+}\tau_qP_q}{1{+}\gamma_\text{e}P_q}\right)\right\rbrace^{\hspace{-0.1cm}+}\right],
\end{align}
and\vspace{-0.3cm}
\begin{align}\label{th1_b}
\mathcal{C}_{\text{s}}^+&{=}\hspace{-0.1cm}\min_{1\leq k\leq K}\max_{\{\tau_q;P_q\}_{q=1}^Q}\sum_{q=0}^Q\Theta_{\tau_q}^{\gamma_k}\nonumber\\
&\qquad\times\underset{\gamma_\text{e},\gamma_k}{\mathbb{E}}\!\left[\left\lbrace\!\log\!\left(\frac{1{+}\gamma_kP_q}{1{+}\gamma_\text{e}P_q}\right)\!\right\rbrace^+\!\Bigg|\tau_q\!\leq\!\gamma_k\!<\!\tau_{q+1}\right],
\end{align}
where $Q{=}2^b$, $\{\tau_q~\!|~\!0{=}\tau_0{<}\tau_1{<}\cdots{<}\tau_Q\}_{q=1}^Q$ represent the reconstruction points describing the support of $\gamma_k$ with $\tau_{Q+1}{=}\infty$ for convenience, $\{P_q\}_{q=1}^Q$ are the power transmission strategies satisfying the average power constraint, and $\displaystyle{\Theta_{\tau_q}^{\gamma_k}{=}\text{Pr}\left[\tau_q{\leq}\gamma_k{<}\tau_{q+1}\right]}$ for all $q\in\{1,{\cdots},Q\}$.
\end{subequations}\vspace{0.2cm}

\textit{Proof:} A detailed proof of Theorem 1 is provided in the following subsection.\vspace{0.2cm}

\noindent It is worth mentioning that the main difference between our bounds is that the transmission scheme, for the achievable secrecy rate, uses the feedback information to adapt both the rate and the power in such a way that the transmission rate is fixed during each fading block. Also, Theorem~1 states that even with 1-bit feedback, sent by each legitimate receiver at the beginning of each fading block, a positive secrecy rate can still be achieved.

\textit{Corollary 1:} The common message secrecy capacity of a block fading broadcast channel with perfect main CSIT, and the average power constraint in \eqref{pavg}, is given by
\begin{align}\label{cr1}
\mathcal{C}_{\text{s}}=\min_{1\leq k\leq K}\max_{P(\gamma_k)}\underset{\gamma_k,\gamma_\text{e}}{\mathbb{E}}\!\left[\left\lbrace\log\!\left(\frac{1{+}\gamma_kP(\gamma_k)}{1{+}\gamma_\text{e}P(\gamma_k)}\right)\right\rbrace^+\right],
\end{align}
with $\mathbb{E} [P(\gamma_k)]\leq P_\text{avg}$. 

\vspace{0.3cm}
\textit{Proof:} Corollary 1 results directly from the expressions of the achievable rate in \eqref{th1_a} and the upper bound in \eqref{th1_b}, by letting $\displaystyle{\Theta_{\tau_q}^{\gamma_k}=\frac{1}{Q}}$ and taking into consideration that as $Q\rightarrow\infty$, the set of reconstruction points, $\{\tau_1,{\cdots},\tau_Q\}$, becomes infinite and each legitimate receiver $\text{R}_k$ is basically forwarding $\gamma_k$ to the transmitter. $\hfill \square$

To the best of our knowledge, this result has not been reported in earlier works. For the special case of single user transmission, the secrecy capacity in corollary 1 coincides with the result in Theorem~2 from reference \cite{gopala01}.
\subsection{Ergodic Capacity Analysis}\label{ECA1}
In this subsection, we establish the obtained results for the ergodic secrecy capacity presented in the previous subsection.
\subsubsection{Proof of Achievability in Theorem 1}\label{PLB_CM} 
Since the transmission is controlled by the feedbacked information, we consider that, during each fading block, if the main channel gain of the receiver with the weakest channel gain falls within the interval $[\tau_q,\tau_{q+1})$, $q\in\{1,{\cdots},Q\}$, the transmitter conveys the codewords at rate $\mathcal{R}_q=\log\left(1{+}\tau_qP_q\right).$ Rate $\mathcal{R}_q$ changes only periodically and is held constant over the duration interval of a fading block. This setup guarantees that when $\gamma_\text{e}{>}\tau_q$, the mutual information between the transmitter and the eavesdropper is upper bounded by $\mathcal{R}_q$. Otherwise, this mutual information will be $\log\left(1{+}\gamma_\text{e}P_q\right)$.

Besides, we adopt a probabilistic transmission model where the communication is constrained by the quality of the legitimate channels. Given the reconstruction points, $\tau_1{<}\tau_2{<}\cdots{<}\tau_Q{<}\tau_{Q+1}{=}\infty$, describing the support of each channel gain $\gamma_k; k\in\{1,\cdots ,K\}$, and since the channel gains of the $K$ receivers are independent, there are $M{=}Q^K$ different states for the received feedback information. Each of these states, $\mathcal{J}_m; m\in\{1,\cdots ,M\},$ represents one subchannel. The transmission scheme consists on transmitting an independent codeword, on each of the $M$ subchannels, with a fixed rate. We define the following rates
$$\displaystyle{\mathcal{R}_\text{s}^-=\sum_{m=1}^M\text{Pr}\left[\mathcal{J}_m\right]\underset{\gamma_\text{e}}{\mathbb{E}}\left[\left\lbrace\log\left(\frac{1{+}\tau_m^\text{min}P_m}{1{+}\gamma_\text{e}P_m}\right)\right\rbrace^+\right],}$$
and $\displaystyle{\mathcal{R}_{\text{e},m}=\underset{\gamma_\text{e}}{\mathbb{E}}\left[\log\left(1{+}\gamma_\text{e}P_m\right)\right],}$
where $\tau_m^\text{min}$ is the quantized channel gain corresponding to the weakest receiver in state $\mathcal{J}_m$ and $P_m$ is the associated power policy satisfying the average power constraint. 

\textit{Codebook Generation:}
We construct $M$ independent codebooks $\mathcal{C}_1$, $\cdots$, $\mathcal{C}_M$, one for each subchannel, constructed similarly to the standard wiretap codes. Each codebook $\mathcal{C}_m$ is a $(n,2^{n\mathcal{R}_\text{s}^-})$ code with $2^{n(\mathcal{R}_\text{s}^-{+}\mathcal{R}_{\text{e},m})}$ codewords randomly partitioned into~$2^{n\mathcal{R}_\text{s}^-}$ bins.

\textit{Encoding and Decoding:}
Given a particular common message $w{\in}\{1,2,{\cdots},2^{n\mathcal{R}_\text{s}^-}\}$, to be transmitted, the encoder selects $M$ codewords, one for each subchannel. More specifically, if the message to be sent is $w$, then for each subchannel $m$, the encoder randomly selects one of the codewords $U_m^n$ from the $w$th bin in $\mathcal{C}_m$. During each fading block, of length $\kappa$, the transmitter experiences one of the events $\mathcal{J}_m$. Depending on the encountered channel state, the transmitter broadcasts $\kappa\mathcal{R}_q$ information bits of $U_m^{n}$ using a Gaussian codebook. By the weak law of large numbers, when the total number of fading blocks $L$ is large, the entire binary sequences are transmitted with high probability. To decode, each legitimate receiver considers the observations corresponding to all $M$ subchannels. And since $\log\left(1{+}\tau_m^\text{min}P_q\right){<}\log\left(1{+}\gamma_kP_q\right)$ is valid for all fading blocks, the receivers can recover all codewords, with high probability, and hence recover message~$w$. 
The expression of $\mathcal{R}_\text{s}^-$ can then be reformulated as
\begin{align}
&\mathcal{R}_\text{s}^-=\sum_{m=1}^M\text{Pr}\left[\mathcal{J}_m\right]\underset{\gamma_\text{e}}{\mathbb{E}}\left[\left\lbrace\log\left(\frac{1{+}\tau_m^\text{min}P_m}{1{+}\gamma_\text{e}P_m}\right)\right\rbrace^+\right]\\
&=\hspace{-0.2cm}\min_{1\leq k\leq K}\sum_{m=1}^M\text{Pr}\left[\mathcal{J}_m\right]\underset{\gamma_\text{e}}{\mathbb{E}}\left[\left\lbrace\log\left(\frac{1{+}\tau_{k,m}P_m}{1{+}\gamma_\text{e}P_m}\right)\right\rbrace^+\right]\label{stp1}\\
&=\hspace{-0.2cm}\min_{1\leq k\leq K}\sum_{m=1}^M\sum_{q=1}^Q\text{Pr}\left[\mathcal{J}_m,\tau_{k,m}{=}\tau_q\right]\underset{\gamma_\text{e}}{\mathbb{E}}\left[\left\lbrace\log\left(\frac{1{+}\tau_qP_q}{1{+}\gamma_\text{e}P_q}\!\right)\!\right\rbrace^{\hspace{-0.1cm}+}\right]\label{stp2}\\
&=\hspace{-0.2cm}\min_{1\leq k\leq K}\sum_{q=1}^Q\text{Pr}\left[\tau_q{\leq}\gamma_k{<}\tau_{q+1}\right]\underset{\gamma_\text{e}}{\mathbb{E}}\!\left[\!\left\lbrace\!\log\!\left(\!\frac{1{+}\tau_qP_q}{1{+}\gamma_\text{e}P_q}\!\right)\!\right\rbrace^{\hspace{-0.1cm}+}\right],\label{stp3}
\end{align}
where \eqref{stp1} results since the logarithm function is monotonic and the sum and the expectation are taking over positive terms, \eqref{stp2} is obtained by noting that $\tau_m^\text{min}\in\{\tau_1,{\cdots},\tau_Q\}$ and applying the total probability theorem, and \eqref{stp3} comes from the fact that $\sum_{m=1}^M\text{Pr}\left[\mathcal{J}_m,\tau_{k,m}{=}\tau_q\right]{=}\text{Pr}\left[\tau_q{\leq}\gamma_k{<}\tau_{q+1}\right]$.

Since each user gets to know the feedback information of the other legitimate receivers, our proof is also valid when the reconstruction points $\{\tau_q\}_{q=1}^Q$ and the transmission strategies $\{P_q\}_{q=1}^Q$, associated with each legitimate receiver, are different. That is, we can choose these quantization parameters to satisfy~\eqref{th1_a}. 

\textit{Secrecy Analysis:} 
We need to prove that the equivocation rate satisfies $\mathcal{R}_\text{e}\geq\mathcal{R}_\text{s}^--\epsilon$. Let $\Gamma^L{=}\left\{\gamma_1^L,\gamma_2^L,{\cdots},\gamma_K^L\right\}$
and $F^L{=}\left\{F_1^L,F_2^L,{\cdots},F_K^L\right\}$, with $F_k(l)\in\{\tau_1,{\cdots},\tau_Q\}$ being the feedback information sent by receiver $k$ in the $l$-th fading block. We have\vspace{-0.1cm}
\begin{align}
&n\mathcal{R}_\text{e}=H(W|Y_\text{e}^n,\gamma_\text{e}^L,\Gamma^L,F^L)\\
&\geq I(W;X^n|Y_\text{e}^n,\gamma_\text{e}^L,\Gamma^L,F^L)\\
&=H(X^n|Y_\text{e}^n\!,\!\gamma_\text{e}^L\!,\!\Gamma^L\!,\!F^L){-}H(X^n|Y_\text{e}^n\!,\!\gamma_\text{e}^L\!,\!\Gamma^L\!,\!F^L\!,\!W).\label{lb1_step0}
\end{align}  \vspace{-0.1cm}
On one hand, we can write
\allowdisplaybreaks
\begin{align}
&H(X^n|Y_\text{e}^n,\gamma_\text{e}^L,\Gamma^L,F^L)\nonumber\\
&=\sum_{l=1}^L\hspace{-0.1cm}\resizebox{.38\textwidth}{!}{$\displaystyle{H(X^\kappa(l)|Y_\text{e}^\kappa(l)\!,\!\gamma_\text{e}(l)\!,\!\gamma_1(l)\!,\!{\cdots}\!,\!\gamma_K(l)\!,\!F_1(l)\!,\!{\cdots}\!,\!F_K(l))}$}\label{lb1_step1}\\
&\geq\!\sum_{l\in\mathcal{D}_L}\hspace{-0.15cm}\resizebox{.38\textwidth}{!}{$\displaystyle{H(X^\kappa(l)|Y_\text{e}^\kappa(l)\!,\!\gamma_\text{e}(l)\!,\!\gamma_1(l)\!,\!{\cdots}\!,\!\gamma_K(l)\!,\!F_1(l)\!,\!{\cdots}\!,\!F_K(l))}$}\label{lb1_step2}\\
&\geq \!\sum_{l\in\mathcal{D}_L}\hspace{-0.15cm}\kappa\!\resizebox{.37\textwidth}{!}{$\displaystyle{\left(\min_{1\leq k\leq K}\sum_{q=1}^Q\Theta_{\tau_q}^{\gamma_k(l)}\left(\mathcal{R}_q{-}\log\left(1{+}\gamma_\text{e}(l)P_q\right)\right){-}\epsilon'\!\right)}$}\\
&=\!\sum_{l=1}^L\hspace{-0.1cm}\kappa\!\left(\!\min_{1\leq k\leq K}\sum_{q=1}^Q\hspace{-0.1cm}\Theta_{\tau_q}^{\gamma_k(l)}\!\left\lbrace \!\mathcal{R}_q\!-\!\log\!\left(1{+}\gamma_\text{e}(l)P_q\right)\!\right\rbrace^{+}\hspace{-0.1cm}\!-\!\epsilon'\!\right)\\
&=n\min_{1\leq k\leq K}\sum_{q=1}^Q\Theta_{\tau_q}^{\gamma_k}~\!\underset{\gamma_\text{e}}{\mathbb{E}}\left[\left\lbrace \mathcal{R}_q{-}\log\left(1{+}\gamma_\text{e}P_q\right)\right\rbrace^+\right]-n\epsilon'\label{lb1_step3}\\
&=n\mathcal{R}_\text{s}^- -n\epsilon',\label{lb1_step3p1}
\end{align}
where \eqref{lb1_step1} results from the memoryless property of the channel and the independence of the $X^\kappa(l)$'s, \eqref{lb1_step2} is obtained by removing all the terms corresponding to the fading blocks $l\not\in\mathcal{D}_L$, with $\mathcal{D}_L=\cup_{k\in\{1,{\cdots},K\}}\left\lbrace l\in\{1,{\cdots},L\}:F_k(l)>h_\text{e}(l)\right\rbrace$, $\displaystyle{\Theta_{\tau_q}^{\gamma_k(l)}{=}\text{Pr}\left[\tau_q{\leq}\gamma_k(l){<}\tau_{q+1}\right]}$, and \eqref{lb1_step3} follows from the ergodicity of the channel as $L\rightarrow\infty$. 

On the other hand, using list decoding argument at the eavesdropper side and applying Fano's inequality~\cite{gopala01}, $\frac{1}{n}H(X^n|Y_\text{e}^n,\gamma_\text{e}^L,\Gamma^L,F^L,W)$ vanishes as $n\rightarrow\infty$ and we can write
\begin{equation}\label{lb1_step4}
H(X^n|Y_\text{e}^n,\gamma_\text{e}^L,\Gamma^L,F^L,W)\leq n\epsilon''.
\end{equation}
Substituting \eqref{lb1_step3p1} and \eqref{lb1_step4} in \eqref{lb1_step0}, we get $\mathcal{R}_\text{e}\geq \mathcal{R}_\text{s}^- -\epsilon$, with $\epsilon=\epsilon'+\epsilon''$, and $\epsilon'$ and $\epsilon''$ are selected to be arbitrarily small. 
This concludes the proof.$\hfill \square$

\subsubsection{Proof of the Upper Bound in Theorem 1}\label{PUB_CM}
To establish the upper bound on the common message secrecy capacity in~(\ref{th1_b}), we start by supposing that the transmitter sends message $w$ to only one legitimate receiver $\text{R}_k$. Using the result in \cite{rezki11}, for single user transmission with limited CSI feedback, the secrecy capacity of our system can be upper bounded as 
\begin{equation}\label{UB_CM_stp1}
\mathcal{C}_{\text{s}}\!\leq\!\hspace{-0.2cm}\max_{\{\tau_q;P_q\}_{q=1}^Q}\sum_{q=0}^Q\Theta_{\tau_q}^{\gamma_k}\underset{\gamma_\text{e},\gamma_k}{\mathbb{E}}\!\left[\!\left\lbrace\!\log\!\left(\!\frac{1{+}\gamma_kP_q}{1{+}\gamma_\text{e}P_q}\!\right)\!\right\rbrace^{\hspace{-0.1cm}+}\!\Bigg|\tau_q\!\leq\!\gamma_k\!<\!\tau_{q+1}\!\right],
\end{equation}
with $\displaystyle{\Theta_{\tau_q}^{\gamma_k}{=}\text{Pr}\left[\tau_q{\leq}\gamma_k{<}\tau_{q+1}\right]}$.

\noindent Since the choice of the receiver to transmit to is arbitrary, we tighten this upper bound by choosing the legitimate receiver $\text{R}_k$ that minimizes this quantity, yielding 
\begin{align}\label{UB_CM_stp2}
\mathcal{C}_{\text{s}}^+&=\min_{1\leq k\leq K}\max_{\{\tau_q;P_q\}_{q=1}^Q}\sum_{q=0}^Q\Theta_{\tau_q}^{\gamma_k}\\
&\qquad\quad\times\underset{\gamma_\text{e},\gamma_k}{\mathbb{E}}\!\left[\!\left\lbrace\!\log\!\left(\!\frac{1{+}\gamma_kP_q}{1{+}\gamma_\text{e}P_q}\!\right)\!\right\rbrace^{\hspace{-0.1cm}+}\!\Bigg|\tau_q\!\leq\!\gamma_k\!<\!\tau_{q+1}\!\right].\qquad \square\nonumber
\end{align}

\section{Broadcasting Independent Messages}\label{BIM}
In this section, we consider the independent messages case when multiple confidential messages are broadcasted to the legitimate receivers in the presence of an eavesdropper. Taking into account the adopted system model, we present an upper and a lower bounds on the ergodic secrecy sum-capacity.
\subsection{Main Results}\label{MR2}
\textit{Theorem 2:} The secrecy sum-capacity, $\widetilde{\mathcal{C}}_{\text{s}}$, of a block fading broadcast channel, with an error free $b$-bits CSI feedback sent by each legitimate receiver, at the beginning of each fading block, is bounded by
\begin{equation}\label{th2}
\widetilde{\mathcal{C}}_{\text{s}}^-\leq \widetilde{\mathcal{C}}_{\text{s}}\leq \widetilde{\mathcal{C}}_{\text{s}}^+,
\end{equation}
such as\vspace{-0.3cm}
\begin{subequations}
\begin{align}\label{th2_a}
\widetilde{\mathcal{C}}_{\text{s}}^-{=}\max_{\{\tau_q;P_q\}_{q=1}^Q}\sum_{q=1}^Q\Theta_{\tau_q}^{\gamma_\text{max}}\hspace{0.2cm}\underset{\gamma_\text{e}}{\mathbb{E}}\left[\left\lbrace\log\left(\frac{1{+}\tau_qP_q}{1{+}\gamma_\text{e}P_q}\right)\right\rbrace^+\right],
\end{align}
and\vspace{-0.3cm}
\begin{align}\label{th2_b}
&\widetilde{\mathcal{C}}_{\text{s}}^+{=}\max_{\{\tau_q;P_q\}_{q=1}^Q}\sum_{q=0}^Q\Theta_{\tau_q}^{\gamma_\text{max}}\nonumber\\
&\times\underset{\gamma_\text{e},\gamma_\text{max}}{\mathbb{E}}\!\left[\left\lbrace\!\log\!\left(\frac{1{+}\gamma_\text{max}P_q}{1{+}\gamma_\text{e}P_q}\right)\!\right\rbrace^+\Bigg|\tau_q\!\leq\!\gamma_\text{max}\!<\!\tau_{q+1}\right],
\end{align}
where $\displaystyle{\gamma_\text{max}{=}\max_{1\leq k\leq K}\gamma_k}$, $Q{=}2^b$, $\{\tau_q~\!|~\!0{=}\tau_0{<}\tau_1{<}\cdots{<}\tau_Q\}_{q=1}^Q$ represent the reconstruction points describing the support of $\gamma_\text{max}$ with $\tau_{Q+1}{=}\infty$ for convenience, $\{P_q\}_{q=1}^Q$ are the power transmission strategies satisfying the average power constraint, and $\Theta_{\tau_q}^{\gamma_\text{max}}{=}\text{Pr}\left[\tau_q{\leq}\gamma_\text{max}{<}\tau_{q+1}\right]$ for all $q\in\{1,{\cdots},Q\}$.
\end{subequations}

\textit{Proof:} A detailed proof of Theorem 2 is provided in the following subsection.
Theorem~2 states that even with 1-bit feedback, sent by the strongest legitimate receiver at the beginning of each fading block, a positive secrecy rate can still be achieved.\vspace{0.2cm}

\textit{Remarks:}
\begin{itemize}
\item The presented results, for both common message and independent messages transmissions, are also valid in the case when multiple non-colluding eavesdroppers conduct the attack. In the case when the eavesdroppers collude, the results can be extended by replacing the term $\gamma_\text{e}$ with the squared norm of the vector of channel gains of the colluding eavesdroppers. 
\item In the analyzed system, we assumed unit variance Gaussian noises at all receiving nodes. The results can be easily extended to a general setup where the noise variances are different.
\end{itemize}

\textit{Corollary 2:} The secrecy sum-capacity of a block fading broadcast channel with perfect main CSIT, and the average power constraint in \eqref{pavg}, is given by
\begin{align}\label{cr2}
\widetilde{\mathcal{C}}_{\text{s}}=\max_{P(\gamma_\text{max})}\underset{\gamma_\text{max},\gamma_\text{e}}{\mathbb{E}}\!\left[\left\lbrace\log\!\left(\frac{1{+}\gamma_\text{max}P(\gamma_\text{max})}{1{+}\gamma_\text{e}P(\gamma_\text{max})}\right)\right\rbrace^+\right],
\end{align}
with $\gamma_\text{max}{=}\max_{1\leq k\leq K}\gamma_k$, and $\mathbb{E}[P(\gamma_\text{max})]\leq P_\text{avg}$. 

\vspace{0.3cm}
\textit{Proof:} Corollary 2 results directly by following a similar reasoning as for the proof of Corollary 1. $\hfill \square$

To the best of our knowledge, this result has not been reported in earlier works. For the special case of single user transmission, the secrecy sum-capacity in corollary 2 coincides with the result in Theorem~2 from reference \cite{gopala01}.

\subsection{Ergodic Capacity Analysis}\label{ECA2}
In this subsection, we establish the obtained results for the ergodic secrecy sum-capacity.
\subsubsection{Proof of Achievability in Theorem 2}\label{PLB_IM} 
The lower bound on the secrecy sum-capacity, presented in \eqref{th2_a}, is achieved using a time division multiplexing scheme that selects periodically one receiver to transmit to. More specifically, we consider that, during each fading block, the source only transmits to the legitimate receiver with the highest $\tau_q$, and if there are more than one, we choose one of them randomly. Since we are transmitting to only one legitimate receiver at a time, the achieving coding scheme consists on using independent standard single user Gaussian wiretap codebooks.  

During each fading block, the transmitter receives~$K$ feedback information about the CSI of the legitimate receivers. Since the channel gains of the $K$ receivers are independent, there are $M{=}Q^K$ different states for the received feedback information, as discussed in the proof of achievability of Theorem 1. Each of these states, $\mathcal{J}_m; m\in\{1,\cdots ,M\},$ represents one subchannel. The transmission scheme consists on sending an independent message, intended for the receiver with the highest $\tau_q$, on each of the $M$ subchannels, with a fixed rate. Let $\tau_m^\text{max}$ be the maximum received feedback information on channel $m$. The overall achievable secrecy sum-rate can be written as
\begin{align}
&\widetilde{\mathcal{R}}_\text{s}^-=\sum_{m=1}^M\text{Pr}[\mathcal{J}_m]~\!\underset{\gamma_\text{e}}{\mathbb{E}}\left[\left\lbrace\log\left(\frac{1{+}\tau_m^\text{max}P(\tau_m^\text{max})}{1{+}\gamma_\text{e}P(\tau_m^\text{max})}\right)\right\rbrace^+\right]\label{Rs_IM_stp1}\\
&=\sum_{q=1}^Q\text{Pr}[\tau_q{\leq}\gamma
_\text{max}{<}\tau_{q+1}]~\!\underset{\gamma_\text{e}}{\mathbb{E}}\left[\left\lbrace\log\left(\frac{1{+}\tau_qP_q}{1{+}\gamma_\text{e}P_q}\right)\right\rbrace^+\right],\label{Rs_IM_stp2}
\end{align} 
where \eqref{Rs_IM_stp1} is obtained by using a Gaussian codebook with power $P(\tau_m^\text{max})$, satisfying the average power constraint, on each subchannel $m$ \cite{gopala01}, and \eqref{Rs_IM_stp2} follows by using the fact that $\tau_m^\text{max}\in\{\tau_1,\cdots ,\tau_Q\}$ and rewriting the summation over these indices. Also, we note that the probability of adapting the transmission with $\tau_q$ corresponds to the probability of having $\tau_q{\leq}\gamma
_\text{max}{<}\tau_{q+1}$, with $\gamma_\text{max}=\max_{1\leq k\leq K}\gamma_k$.
Maximizing over the main channel gain reconstruction points $\tau_q$ and the associated power transmission strategies $P_q$, for each $q\in\{1,\cdots ,Q\}$, concludes the proof.$\hfill \square$
\subsubsection{Proof of the Upper Bound in Theorem 2}\label{PUB_IM}
To prove that $\widetilde{\mathcal{C}}_{\text{s}}^+$, presented in \eqref{th2_b}, is an upper bound on the secrecy sum-capacity, we consider a new genie-aided channel whose capacity upper bounds the capacity of the $K$-receivers channel with limited CSI feedback. The new channel has only one receiver that observes the output of the strongest main channel. The output signal of the genie-aided receiver is given by $Y_\text{max}(t)=h_\text{max}(i)X(t)+v(t)$, at time instant $t$, with $h_\text{max}$ being the channel gain of the best legitimate channel, i.e., $|h_\text{max}|^2{=}\gamma_\text{max}$ and $\gamma_\text{max}{=}\max_{1\leq k\leq K}\gamma_k$. 
Let $\tau_q; q\in\{1,{\cdots},Q\}$ be the feedback information sent by the new receiver to the transmitter about its channel gain, i.e., $\tau_q$ is feedbacked when $\tau_q{\leq}\gamma_\text{max}{<}\tau_{q+1}$. First, we need to prove that the secrecy capacity of this new channel upper bounds the secrecy sum-capacity of the $K$-receivers channel with limited CSI. To this end, it is sufficient to show that if a secrecy rate point $(\mathcal{R}_1,\mathcal{R}_2,{\cdots},\mathcal{R}_K)$ is achievable on the $K$-receivers channel with limited CSI feedback, then, a secrecy sum-rate $\sum_{k=1}^K\mathcal{R}_k$ is achievable on the new channel.

Let $(W_1,\!W_2,\!{\cdots},\!W_K)$ be the independent transmitted messages corresponding to the rates $(\mathcal{R}_1,\!\mathcal{R}_2,\!{\cdots},\!\mathcal{R}_K)$, and $(\hat{W}_1,\hat{W}_2,{\cdots},\hat{W}_K)$ the decoded messages. Thus, for any $\epsilon{>}0$ and $n$ large enough, there exists a code of length $n$ such that $\text{Pr}[\hat{W}_k\!\neq\!W_k]\!\leq\!\epsilon$ at each of the $K$ receivers, and
\begin{equation}
\frac{1}{n}H\!(W_k|W_1\!,\!\cdots\!,W_{k\!-\!1}\!,\!W_{k\!+\!1}\!,\!\cdots\!,\!W_K\!,\!Y_\text{e}^n\!,\!\gamma_\text{e}^L\!,\!\!F^L)\!\geq\!\mathcal{R}_k\!-\!\epsilon ,\label{sc23}
\end{equation}
with $F^L{=}\{F_1^L,F_2^L,{\cdots},F_K^L\}$, and $F_k(l)~\!{\in}~\!\{\tau_1,{\cdots},\tau_Q\}$ is the feedback information sent by receiver $k$ in the $l$-th fading block. Now, we consider the transmission of message $W{=}(W_1\!,\!W_2\!,\!\cdots\!,\!W_K)$ to the genie-aided receiver~using the same encoding scheme as for the $K$-receivers case. Adopting a decoding scheme similar to the one used at each~of the $K$ legitimate receivers, it is clear that the genie-aided~receiver can decode message $W$ with a negligible probability of error, i.e., $\text{Pr}(\hat{W}\!\neq\!W)\!\leq\!\epsilon$. For the secrecy condition, we have
\begin{align}
&\frac{1}{n}H\!(W|Y_\text{e}^n\!,\!\gamma_\text{e}^L\!,\!\gamma_\text{max}^L,\!F_\text{max}^L)\nonumber\\
&=\frac{1}{n}H\!(W_1\!,\!W_2\!,\!\cdots\!,\!W_K|Y_\text{e}^n\!,\!\gamma_\text{e}^L\!,\!\gamma_\text{max}^L,\!F_\text{max}^L)\\
&\geq\hspace{-0.1cm}\sum_{k=1}^K\!\frac{1}{n}H\!(W_k|W_1,\!\cdots\!,W_{k\!-\!1}\!,\!W_{k\!+\!1}\!,\!\cdots\!,\!W_K\!,\!Y_\text{e}^n\!,\!\gamma_\text{e}^L\!,\!\gamma_\text{max}^L,\!F_\text{max}^L)\\
&\geq\hspace{-0.1cm}\sum_{k=1}^K\!\frac{1}{n}H\!(W_k|W_1,\!\cdots\!,W_{k\!-\!1}\!,\!W_{k\!+\!1}\!,\!\cdots\!,\!W_K\!,\!Y_\text{e}^n\!,\!\gamma_\text{e}^L\!,\!\gamma_\text{max}^L,\!F^L)\label{UB_IM_v0}
\end{align}\vspace{-0.4cm}
\begin{equation}
\geq\hspace{-0.1cm}\sum_{k=1}^K\mathcal{R}_k{-}K\epsilon ,\label{UB_IM_v01}\hspace{5.5cm}
\end{equation}
where $F_\text{max}^L{=}\{F_\text{max}(1),{\cdots},F_\text{max}(L)\}$ and $F_\text{max}(l)$ is the feedback information sent by the genie-aided receiver in the $l$-th fading block, \eqref{UB_IM_v0} follows from the fact that $F_\text{max}{\in}\{F_1,{\cdots},F_K\}$ and that conditioning reduces the entropy, and where \eqref{UB_IM_v01} follows from the secrecy constraint \eqref{sc23}. 

Now, we need to prove that $\widetilde{\mathcal{C}}_\text{s}^+$ upper bounds the secrecy capacity of the genie-aided channel. Let $\widetilde{\mathcal{R}}_\text{e}$ be the equivocation rate of the new channel. We have
\allowdisplaybreaks
\begin{align}
&n\widetilde{\mathcal{R}}_\text{e}=H(W|Y_\text{e}^n,\gamma_\text{e}^L,\gamma_\text{max}^L,F_\text{max}^L)\label{Re_step1}\\
&=I(W\!;\!Y_\text{max}^n|Y_\text{e}^n\!,\!\gamma_\text{e}^L\!,\!\gamma_\text{max}^L\!,\!F_\text{max}^L){+}H(W|Y_\text{max}^n\!,\!Y_\text{e}^n\!,\!\gamma_\text{e}^L\!,\!\gamma_\text{max}^L\!,\!F_\text{max}^L)\\
&\leq I(W;Y_\text{max}^n|Y_\text{e}^n,\gamma_\text{e}^L,\gamma_\text{max}^L,F_\text{max}^L){+} n\epsilon \label{Re_step2}\\
&=\sum_{l=1}^L\sum_{k=1}^\kappa\hspace{-0.1cm}I(W\!;\!Y_\text{max}(l,k)|Y_\text{e}^n\!,\!\gamma_\text{e}^L\!,\!\gamma_\text{max}^L,F_\text{max}^L,Y_\text{max}^{\kappa(l\!-\!1)\!+\!(k\!-\!1)}){+}n\epsilon \\
&=\sum_{l=1}^L\sum_{k=1}^\kappa H(Y_\text{max}(l,k)|Y_\text{e}^n,\gamma_\text{e}^L,\gamma_\text{max}^L,F_\text{max}^L,Y_\text{max}^{\kappa(l\!-\!1)\!+\!(k\!-\!1)})\nonumber\\
&-H(Y_\text{max}(l,k)|W\!,\!Y_\text{e}^n\!,\!\gamma_\text{e}^L\!,\!\gamma_\text{max}^L\!,\!F_\text{max}^L\!,\!Y_\text{max}^{\kappa(l\!-\!1)\!+\!(k\!-\!1)}){+} n\epsilon \\
&\leq \sum_{l=1}^L\sum_{k=1}^\kappa H(Y_\text{max}(l,k)|Y_\text{e}(l,k),\gamma_\text{e}(l),\gamma_\text{max}(l),F_\text{max}^l)\\
&-H(Y_\text{max}(l,k)|W\!,\!X(l,k)\!,\!Y_\text{e}^n,\gamma_\text{e}^L\!,\!\gamma_\text{max}^L\!,\!F_\text{max}^L\!,\!Y_\text{max}^{\kappa(l\!-\!1)\!+\!(k\!-\!1)}){+} n\epsilon\nonumber\\
&=\sum_{l=1}^L\sum_{k=1}^\kappa H(Y_\text{max}(l,k)|Y_\text{e}(l,k),\gamma_\text{e}(l),\gamma_\text{max}(l),F_\text{max}^l)\\
&-H(Y_\text{max}(l,k)|X(l,k),Y_\text{e}(l,k),\gamma_\text{e}(l),\gamma_\text{max}(l),F_\text{max}^l){+} n\epsilon\nonumber\\
&= \sum_{l=1}^L\sum_{k=1}^\kappa\hspace{-0.1cm}I(X(l\!,\!k);\!Y_\text{max}(l\!,\!k)|Y_\text{e}(l,k)\!,\!\gamma_\text{e}(l)\!,\!\gamma_\text{max}(l)\!,\!F_\text{max}^l)\!+\!n\epsilon
\end{align}\vspace{-0.8cm}
\begin{align}
&\leq \sum_{l=1}^L\sum_{k=1}^\kappa \left\lbrace I(X(l,k);Y_\text{max}(l,k)|\gamma_\text{max}(l),F_\text{max}^l)\right.\hspace{2cm}\nonumber\\
&\left.\hspace{1.6cm}-I(X(l,k);Y_\text{e}(l,k)|\gamma_\text{e}(l),F_\text{max}^l)\right\rbrace^+ {+} n\epsilon \label{Re_step3}\\
&= \sum_{l=1}^L\kappa \left\lbrace I(X(l);Y_\text{max}(l)|\gamma_\text{max}(l),F_\text{max}^l)\right.\nonumber\\
&\left.\hspace{1.6cm}-I(X(l);Y_\text{e}(l)|\gamma_\text{e}(l),F_\text{max}^l)\right\rbrace^+ {+} n\epsilon ,\label{Re_step3p1}
\end{align}
where inequality \eqref{Re_step2} follows from the fact that $H(W|Y_\text{max}^n,Y_\text{e}^n,\gamma_\text{e}^L,\gamma_\text{max}^L,F_\text{max}^L){\leq}H(W|Y_\text{max}^n,\gamma_\text{max}^L,F_\text{max}^L),$ and Fano's inequality $H(W|Y_\text{max}^n,\gamma_\text{max}^L,F_\text{max}^L){\leq}n\epsilon ,$ and \eqref{Re_step3} holds true by selecting the appropriate value for the noise correlation to form the Markov chain $X(l){\rightarrow}Y_\text{max}(l){\rightarrow}Y_\text{e}(l)$ if $\gamma_\text{max}(l){>}\gamma_\text{e}(l)$ or $X(l){\rightarrow}Y_\text{e}(l){\rightarrow}Y_\text{max}(l)$ if $\gamma_\text{max}(l){\leq}\gamma_\text{e}(l)$, as explained in \cite{liang01}.

The right-hand side of \eqref{Re_step3p1} is maximized by a Gaussian input. That is, taking $X(l){\sim}\mathcal{CN}\left(0,\omega_l^{1/2}(F_\text{max}^l)\right)$, with the power policy $\omega_l(F_\text{max}^l)$ satisfying the average power constraint, we can write
\begin{align}
&n\widetilde{\mathcal{R}}_\text{e}\leq\kappa\sum_{l=1}^L\mathbb{E}\!\left[\!\left\lbrace\log\left(\frac{1{+}\gamma_\text{max}(l)\omega_l(F_\text{max}^l)}{1{+}\gamma_\text{e}(l)\omega_l(F_\text{max}^l)}\right)\right\rbrace^+\right]{+}n\epsilon \\
&{=}\kappa\sum_{l=1}^L\mathbb{E}\!\left[\!\mathbb{E}\!\resizebox{.35\textwidth}{!}{$\displaystyle{\left[\!\left\lbrace\!\log\!\left(\!\frac{1{+}\gamma_\text{max}(l)\omega_l(F_\text{max}^l)}{1{+}\gamma_\text{e}(l)\omega_l(F_\text{max}^l)}\!\right)\!\right\rbrace^{\hspace{-0.15cm}+}\!\bigg|F_\text{max}(l),\gamma_\text{max}(l),\gamma_\text{e}(l)\right]}$}\right]\hspace{-0.15cm}{+}n\epsilon \\
&{\leq}\kappa\sum_{l=1}^L\mathbb{E}\!\left[\resizebox{.36\textwidth}{!}{$\displaystyle{\!\left\lbrace\log\!\left(\!\frac{1{+}\gamma_\text{max}(l)\mathbb{E}\!\left[\omega_l\scriptstyle{(F_\text{max}^l)|F_\text{max}(l),\gamma_\text{max}(l),\gamma_\text{e}(l)}\right]}{1{+}\gamma_\text{e}(l)\mathbb{E}\!\left[\omega_l\scriptstyle{(F_\text{max}^l)|F_\text{max}(l),\gamma_\text{max}(l),\gamma_\text{e}(l)}\right]}\right)\!\right\rbrace^{\hspace{-0.15cm}+}}$}\right]\hspace{-0.15cm}{+}n\epsilon \label{Re_step4} \\
&=\kappa\sum_{l=1}^L\mathbb{E}\!\left[\left\lbrace\log\left(\frac{1{+}\gamma_\text{max}(l)\Omega_l(F_\text{max}(l))}{1{+}\gamma_\text{e}(l)\Omega_l(F_\text{max}(l))}\right)\!\right\rbrace^{\hspace{-0.15cm}+}\right]\hspace{-0.1cm}{+}n\epsilon \label{Re_step5} \\
&=\kappa\sum_{l=1}^L\mathbb{E}\left[\left\lbrace\log\left(\frac{1{+}\gamma_\text{max}\Omega_l(F_\text{max})}{1{+}\gamma_\text{e}\Omega_l(F_\text{max})}\right)\right\rbrace^+\right]\!+\!n\epsilon ,\label{Re_step6}
\end{align}
where \eqref{Re_step4} is obtained using Jensen's inequality, $\Omega_l(F_\text{max}(l))$ in \eqref{Re_step5} is defined as $\displaystyle{\Omega_l(F_\text{max}(l)){=}\mathbb{E}\left[\omega_l(F_\text{max}^l)|F_\text{max}(l),\gamma_\text{max}(l),\gamma_\text{e}(l)\right],}$ and where \eqref{Re_step6} follows from the ergodicity and the stationarity of the channel gains, i.e., the expectation in \eqref{Re_step5} does not depend on the block fading index. Thus, we have
\begin{align}
&\widetilde{\mathcal{R}}_\text{e}\leq\frac{1}{L}\sum_{l=1}^L\mathbb{E}\!\left[\left\lbrace\log\left(\frac{1{+}\gamma_\text{max}\Omega_l(F_\text{max})}{1{+}\gamma_\text{e}\Omega_l(F_\text{max})}\right)\right\rbrace^+\right]\!+\!\epsilon\\
&\leq\mathbb{E}\left[\left\lbrace\log\left(\frac{1{+}\gamma_\text{max}\Omega(F_\text{max})}{1{+}\gamma_\text{e}\Omega(F_\text{max})}\right)\right\rbrace^+\right]+\epsilon ,\label{Re_step7}
\end{align}
where \eqref{Re_step7} comes from applying Jensen's inequality once again, with $\displaystyle{\Omega(F_\text{max}){=}\frac{1}{L}\sum_{l=1}^L\Omega_l(F_\text{max}).}$
Maximizing over the main channel gain reconstruction points $\tau_q$ and the associated power transmission strategies $P_q$, for each $q\in\{1,\cdots ,Q\}$, concludes the proof.$\hfill \square$
\section{Numerical Results}\label{NR}
\begin{figure}[t]
\psfrag{}[l][l][1]{}
\psfrag{rate}[l][l][1.3]{\hspace{-1cm}Secrecy Rate (npcu)}
\psfrag{snr}[l][l][1.3]{SNR (dB)}
\psfrag{Secrecy Capacity with Perfect Main CSI--------}[l][l][1.2]{Secrecy Capacity $\mathcal{C}_\text{s}$ with Perfect Main CSI}
\psfrag{Secrecy Rate with 6-bits CSI Feedback}[l][l][1.2]{Secrecy Rate $\mathcal{C}_\text{s}^-$ with 6-bits CSI Feedback}
\psfrag{Secrecy Rate with 4-bits CSI Feedback}[l][l][1.2]{Secrecy Rate $\mathcal{C}_\text{s}^-$ with 4-bits CSI Feedback}
\psfrag{Secrecy Rate with 2-bits CSI Feedback}[l][l][1.2]{Secrecy Rate $\mathcal{C}_\text{s}^-$ with 2-bits CSI Feedback}
\psfrag{Secrecy Rate with 1-bit CSI Feedback}[l][l][1.2]{Secrecy Rate $\mathcal{C}_\text{s}^-$ with 1-bit CSI Feedback}
\vspace{-0.7cm}
\begin{center}%
\scalebox{0.55}{\includegraphics{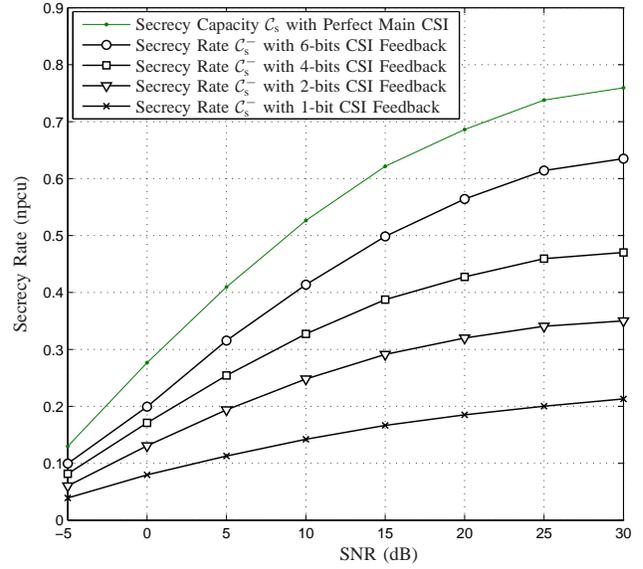}}
\end{center}
\vspace{-0.7cm}
\caption{Common message secrecy rate, for Rayleigh fading channels, with $K{=}3$ and various $b$-bits feedback, $b{=}6,4,2,1.$}
\label{fig:fig1}
\end{figure}

In this section, we provide selected simulation results for the case of Rayleigh fading channels with $\mathbb{E}\left[\gamma_\text{e}\right]{=}\mathbb{E}\left[\gamma_k\right]{=}1$; $k~\!{\in}~\!\{1,{\cdots},K\}$. Figure~\ref{fig:fig1} illustrates the common message achievable secrecy rate $\mathcal{C}_\text{s}^-$, in nats per channel use (npcu), with $K{=}3$ and various $b$-bits feedback, $b{=}6,4,2,1.$ The secrecy capacity $\mathcal{C}_\text{s}$, from Corollary 1, is also presented as a benchmark. It represents the secrecy capacity with full main CSI at the transmitter. We can see that, as the capacity of the feedback link grows, i.e., the number of bits $b$ increases, the achievable rate grows toward the secrecy capacity $\mathcal{C}_\text{s}$. 

\begin{figure}[ht]
\psfrag{}[l][l][1]{}
\psfrag{rate}[l][l][1.3]{\hspace{-1.3cm}Secrecy Sum-Rate (npcu)}
\psfrag{snr}[l][l][1.3]{SNR (dB)}
\psfrag{Secrecy Capacity with Perfect Main CSI and k=3--------}[l][l][1.15]{Secrecy Capacity $\widetilde{\mathcal{C}}_\text{s}$ with Perfect Main CSI and $K{=}3$}
\psfrag{Secrecy Capacity with Perfect Main CSI and k=6}[l][l][1.15]{Secrecy Capacity $\widetilde{\mathcal{C}}_\text{s}$ with Perfect Main CSI and $K{=}6$}
\psfrag{Secrecy Rate with 4-bits CSI Feedback and k=3}[l][l][1.15]{Secrecy Rate $\widetilde{\mathcal{C}}_\text{s}^-$ with 4-bits CSI Feedback and $K{=}3$}
\psfrag{Secrecy Rate with 4-bits CSI Feedback and k=6}[l][l][1.15]{Secrecy Rate $\widetilde{\mathcal{C}}_\text{s}^-$ with 4-bits CSI Feedback and $K{=}6$}
\vspace{-0.55cm}
\begin{center}%
\scalebox{0.55}{\includegraphics{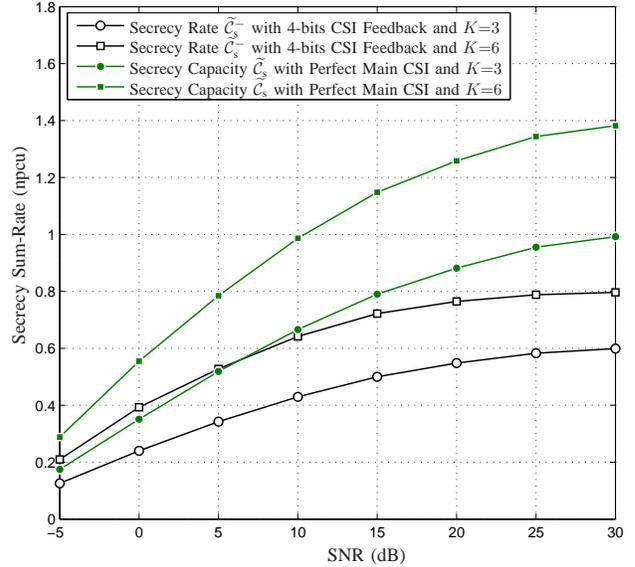}}
\end{center}
\vspace{-0.7cm}
\caption{Independent messages secrecy sum-rate, for Rayleigh fading channels, with $b{=}4$ and two different values of the total number of legitimate receivers, $K{=}3$ and $K{=}6$.}\vspace{-0.4cm}
\label{fig:fig2}
\end{figure}

The independent messages transmission case is illustrated in figure~\ref{fig:fig2}. Two scenarios are considered; the transmission of three independent messages to three legitimate receivers, $K{=}3$, and the transmission of six independent messages with $K{=}6$. Both the achievable secrecy sum-rate $\widetilde{\mathcal{C}}_\text{s}^-$, with 4-bits CSI feedback, and the secrecy capacity $\widetilde{\mathcal{C}}_\text{s}$, with perfect main CSI, are depicted. The curves are presented in npcu. From this figure, we can see that the secrecy throughput of the system, when broadcasting multiple messages, increases with the number of legitimate receivers $K$.

\section{Conclusion}\label{conclusion}
In this work, we analyzed the ergodic secrecy capacity of a broadcast block-fading wiretap channel with limited CSI feedback. Assuming full CSI on the receivers' side and an average power constraint at the transmitter, we presented lower and upper bounds on the ergodic secrecy capacity and the sum-capacity when the feedback link is limited to $b$ bits per fading block. The feedback bits are provided to the transmitter by each legitimate receiver, at the beginning of each coherence block, through error-free public links with limited capacity. The obtained results show that the secrecy rate when broadcasting a common message is limited by the legitimate receiver having, on average, the worst main channel link, i.e., the legitimate receiver with the lowest average SNR. For the independent messages case, we proved that the achievable secrecy sum-rate is contingent on the legitimate receiver with the best instantaneous channel link. Furthermore, we showed that the presented bounds coincide, asymptotically, as the capacity of the feedback links become large, i.e. $b\rightarrow\infty$; hence, fully characterizing the secrecy capacity and the sum-capacity in this case. As an extension of this work, it would be of interest to examine the behavior of the secrecy capacity bounds in the low and the high SNR regimes. It would also be interesting to look at the scaling laws of the system as the number of legitimate receivers increases.

\bibliographystyle{IEEEtran}
\bibliography{Ref1,Ref2}

\balance  

\end{document}